\documentclass[a4paper,11pt]{article}
\pdfoutput=1 

\usepackage{jcappub} 
\usepackage[T1]{fontenc} 

\usepackage{comment}
\usepackage{color}
\usepackage{multirow}
\usepackage{multicol}  
\usepackage{bm} 
\usepackage{cprotect} 
\usepackage{diagbox}
\usepackage{array,ragged2e}

\newcommand{\vek}{\mathbf}
\newcommand{\D}{\vek D}
\newcommand{\M}{\vek M}
\newcommand{\matC}{\mathbf C}
\newcommand{\zobs}{z}
\newcommand{\zcol}{z_\mathrm{col}}
\newcommand{\bl}{b^{\mathrm{L}}}
\newcommand{\be}{b}
\newcommand{\dm}{\delta_\mathrm{m}}
\newcommand{\dcb}{\delta_\mathrm{cb}}
\newcommand{\dnu}{\delta_\mathrm{\nu}}
\newcommand{\dg}{\delta_\mathrm{g}}
\newcommand{\fnu}{f_\mathrm{\nu}}
\newcommand{\fcb}{f_\mathrm{cb}}
\newcommand{\dcrit}{\delta_\mathrm{crit}}
\newcommand{\dcbl}{\delta_\mathrm{cb, L}}
\newcommand{\nisdb}{\mathcal{T}}

\newcommand{\m}{\mathrm{m}}
\newcommand{\cb}{\mathrm{cb}}
\newcommand{\n}{\mathrm{\nu}}
\newcommand{\g}{\mathrm{g}}
\newcommand{\cbnu}{\mathrm{cb, \nu}}
\newcommand{\gm}{\mathrm{g, m}}
\newcommand{\kap}{\mathrm{\kappa}}
\newcommand{\gD}{\mathrm{g, D}}

\newcommand{\kapI}{\mathrm{\kappa, I}}

\newcommand{\plin}{P^{\mathrm{lin}}}
\newcommand{\pnl}{P^{\mathrm{nl}}}

\newcommand{\baseline}{$\bar{b}^i \nisdb (k,z) \dm(k,z)$}
\newcommand{\bmat}{$\bar{b}^i \dm(k,z)$}
\newcommand{\bcb}{$\bar{b}^i \dcb(k,z)$}
\defcitealias{LoVerde_bias}{L14}
\newcolumntype{P}[1]{>{\RaggedRight\arraybackslash}p{#1}}

\title{Modeling Neutrino-Induced Scale-Dependent Galaxy Clustering for Photometric Galaxy Surveys}

\author[a]{P. Rogozenski}
\author[b,a]{E. Krause}
\author[c,d]{V. Miranda}

\affiliation[a]{Department of Physics, University of Arizona \\ Tucson, Arizona, 85721, USA}
\affiliation[b]{Steward Observatory, Department of Astronomy, University of Arizona \\ Tucson, Arizona, 85721, USA}
\affiliation[c]{C. N. Yang Institute for Theoretical Physics, Stony Brook University \\ Stony Brook, NY, 11794, USA}
\affiliation[d]{Department of Physics \& Astronomy, Stony Brook University \\ Stony Brook, NY 11794, USA}

\emailAdd{paulrogozenski@arizona.edu}
\emailAdd{krausee@arizona.edu}
\emailAdd{vivan.miranda@stonybrook.edu}

\abstract{The increasing statistical precision of photometric redshift surveys requires improved accuracy of theoretical predictions for large-scale structure observables to obtain unbiased cosmological constraints. In $\Lambda$CDM cosmologies, massive neutrinos stream freely at small cosmological scales, suppressing the small-scale power spectrum. In massive neutrino cosmologies, galaxy bias modeling needs to accurately relate the scale-dependent growth of the underlying matter field to observed galaxy clustering statistics. In this work, we implement a computationally efficient approximation of the neutrino-induced scale-dependent bias (NISDB). Through simulated likelihood analyses of Dark Energy Survey Year 3 (DESY3) and Legacy Survey of Space and Time Year 1 (LSSTY1) synthetic data that contain an appreciable NISDB, we examine the impact of linear galaxy bias and neutrino mass modeling choices on cosmological parameter inference. We find model misspecification of the NISDB approximation and neutrino mass models to decrease the constraining power of photometric galaxy surveys and cause parameter biases in the cosmological interpretation of future surveys. We quantify these biases and devise mitigation strategies.}

\begin{document}
\maketitle
\flushbottom


\section{Introduction}

The Standard Model of particle physics initially predicted three flavors of massless neutrinos. However, experiments indicate that neutrinos oscillate between flavor eigenstates and thus carry non-zero mass \cite{Sanchez_2003, Ashie_2005, Abdurashitov_2002, Altmann_2005, Fukuda_2002, Ahmad_2002}. These neutrino-oscillation experiments have measured the mass difference between at least two of the three neutrino mass eigenstates. One can rule out models for the relative ordering of the mass differences, or the neutrino mass hierarchy, by constraining the total neutrino mass, $\sum m_\n$, to be less than the minimum allowed $\sum m_\n$ of a given mass hierarchy. To date, the most stringent upper-bound of the total neutrino mass through ground-based experiments was determined by the KATRIN beta-decay experiment to be $\sum m_\n < 0.8$eV at 90\% confidence \citep{katrin}. 

The tightest bounds on $\sum m_\n$, though, are currently found through cosmological observations, since massive neutrinos impact the expansion history, energy density, and structure growth of the universe (see \citep{whitepaper} for a summary of recent cosmological constraints). Massive neutrinos, being relativistic in the early universe, contribute to the radiation energy density. In turn, the amplitude and position of Cosmic Microwave Background (CMB) angular temperature power spectrum peaks \citep{Archidiacono_2017} as well as the Baryon-Acoustic Oscillation (BAO) peak \citep{Follin_2015, peloso2015effect} are influenced by the presence of massive neutrinos.

The expansion of the universe eventually cools massive neutrinos to be non-relativistic when their temperature is less than their rest-mass energy.\footnote{For example, a 0.2eV massive neutrino will become non-relativistic around a redshift of 400 given current CMB temperature constraints \citep{planck_18}.} Non-relativistic neutrinos in the late-time universe maintain a large thermal motion and free-stream at small physical scales (for reviews, see \citep{Lesgourgues_2012, Brinckmann_2019}). Neutrino free-streaming suppresses the small-scale matter power spectrum, which is most directly observable through galaxy weak lensing. The small-scale matter power suppression is indirectly observable through damped correlations in void clustering \citep{Kreisch_2019, Banerjee_2020} and galaxy clustering statistics \citep{Hu_1998, Lesgourgues_2012, green2021cosmological}. At large physical scales, massive neutrinos are coherent with Cold Dark Matter (CDM) and baryon density fluctuations (hereafter denoted as $\cb$) and induce a relative increase in galaxy clustering correlation statistics relative to small scales \citep{LoVerde_2014, Villaescusa_Navarro_2014, Banerjee_2016}. 

Cosmological data analyses of the CMB by the Planck collaboration constrain $\sum m_\n < 0.241$eV at 95\% confidence, while omitting high-multipole CMB polarization data constrains $\sum m_\n < 0.54$eV at 95\% confidence \citep{planck_18}. Analyzing spectroscopic galaxy clustering data from eBOSS with Planck CMB data places $\sum m_\n < 0.129$eV at 95\% confidence \citep{Alam_2021}. Combining eBOSS and Planck data with angular galaxy clustering and lensing from the Dark Energy Survey Year 3 (DESY3) data finds $\sum m_\n < 0.13$eV at 95\% confidence \citep{y3-cosmology}, with the constraining power primarily coming from eBOSS \citep{Alam_2021} and Planck CMB \citep{planck_18} data. With the increased statistical power of future photometric surveys, the next generation of angular clustering and weak lensing measurements are expected to contribute significantly to cosmological constraints on the sum of the neutrino masses, particularly in extended cosmological models (e.g. \citep{PhysRevD.97.123544}). To obtain unbiased constraints from these precise large-scale structure measurements, accurate models of non-linear structure growth in the presence of massive neutrinos are required. 

The non-linear evolution of the (dark) matter distribution in cosmologies with massive neutrinos can be obtained through N-body simulations \citep{Bird_2012, Ali_Ha_moud_2012, Castorina_2015, Banerjee_2016, Adamek_2016, Wright_2017, Rizzo_2017, Villaescusa_Navarro_2018, n-body-loverde, quijote}. For use in cosmological inference, summary statistics, like power spectra, are then measured from simulations and interpolated in (cosmological) parameter space using fitting functions or emulators (see \citep{adamek2022euclid, derose2023aemulus} for recent comparisons of non-linear power spectrum models). The relation of the observed distribution of galaxies to the underlying matter field is described by galaxy bias, which on quasi-linear scales can be described by perturbative expansions with bias coefficients (see \citep{bias-bible} for a recent review). Perturbative bias models extend the range of model validity beyond linear scales and increase the inference precision of the total neutrino mass \citep{Aviles_2020, Modi_2017, potato-plots-2, n-body-worry}.

However, the scale-dependent growth and corresponding reduction in clustering due to massive neutrinos are not captured by perturbative bias models. At the linear galaxy bias level, the scale-dependent growth induced by massive neutrinos introduces a scale-dependence of linear galaxy bias. Predictions for this linear scale-dependent halo bias in the peak-background split context are found by solving the multi-fluid spherical collapse as a function of a given halo mass \citep{LoVerde_bias, RelicFast}, which can then be related to galaxy samples. Utilizing improved modeling of the halo bias in the presence of massive neutrinos have been shown to strengthen constraints on the total neutrino mass and aid in determining the neutrino mass hierarchy \citep{Hernandez, RelicCLASS, RelicCLASS-S4, Ballardinin-Maartens, Tanidis_2020}. However, such models introduce additional model parameters of the non-linear galaxy-halo-connection (see \citep{Wechsler_2018} for a recent review).

Future surveys, like the Legacy Survey of Space and Time\footnote{https://lsstdesc.org/} (LSST), will push deeper in redshift, larger in sky area, and higher in galaxy density to considerably advance our understanding of the cosmos and structure formation. The increasing precision of future surveys then will require accurate models that describe non-linear effects, including those due to massive neutrinos, that do not degrade cosmological parameter inference through additional model complexities. To progress towards this goal, we present an analytic approximation of the Neutrino-Induced Scale-Dependent Bias (NISDB) based on the foundational work of \cite{LoVerde_bias} (hereafter denoted \citetalias{LoVerde_bias}). We show that this approximation models the shape of the linear galaxy bias to good approximation and does not introduce additional galaxy-halo-connection nuisance parameters to photometric redshift galaxy survey likelihood evaluations. We investigate the impact of assumed linear galaxy bias and neutrino mass models on cosmological parameter inference by contaminating DESY3 and LSST Year 1 (LSSTY1) synthetic data with a substantial NISDB in the first full-likelihood 3x2pt analysis of its kind. We additionally examine the robustness of systematic bias criteria of the aforementioned galaxy clustering modeling choices on cosmological parameter inference and fits to the fiducial data. We choose our fiducial total neutrino mass value to be 0.5eV split between 3 degenerate-mass neutrinos to avoid prior-boundary effects and isolate the impact modeling choices have on cosmological parameter inference, which may be relevant even at lower neutrino masses for more stringent analyses combining LSST with next-generation CMB data. 

This paper is structured as follows: In section \ref{sec::theory}, we briefly review the theoretical background of neutrino-induced scale-dependent clustering, deriving a simple approximation of the scale-dependent linear galaxy bias induced by neutrinos. In section \ref{sec::pipeline}, we detail our synthetic likelihood analysis pipeline, fiducial cosmology, and systematic parameters for DESY3- and LSSTY1-like analyses. We then investigate parameter constraints, degeneracies, and biases when changing the linear galaxy bias model and/or the neutrino mass model used to fit the fiducial synthetic data in section \ref{sec::results}. We present our conclusions in section \ref{sec::conclusion}.

\section{Theory and Implementation}
\label{sec::theory}

In this section, we motivate approximations to the NISDB derived in \citetalias{LoVerde_bias} at the linear galaxy bias level for efficient evaluation in cosmological inference. We find our form to be a good approximation of the NISDB and validate our approximate form against \texttt{RelicFAST}\footnote{https://github.com/JulianBMunoz/RelicFAST} calculations \citep{RelicFast}. 

A linear (Eulerian) galaxy bias relation, $b(z)$, relates the density contrast of galaxies, $\dg$, to the underlying total matter field fluctuations, $\dm$, 
\begin{equation}
\label{eqn::lin-gal-bias-def}
    \dg(\vek{x},z) \approx b(z) \dm(\vek{x},z),
\end{equation}
where x is a real-space coordinate and z denotes redshift.

The Peak-Background Split (PBS) formalism \citep{1974ApJ...187..425P, Sheth_1999, Schmidt_2013} shows that the presence of a long-wavelength mode in the dark matter + baryon distribution, $\dcbl$, affects the effective critical density threshold of halo collapse, $\delta_\mathrm{crit}$, and provides an expression for the linear Lagrangian bias, $\bl$. Given a halo mass function, $n(M)$, we can relate the observed halo density contrast to $\dcbl$ through the linear halo bias, written as
\begin{equation}
\label{eqn::lagrangian_bias}
\be(M,k,z) = 1+  \bl(M,k,z) = 1+\frac{\partial \ln n(M,z)}{\partial \dcrit(k,z)} \frac{d \dcrit(k,z)}{d \dcbl (k,z)},
\end{equation}
where $k$ is a Fourier mode, and $\be = 1+  \bl$ is the relation between linear Eulerian and Lagrangian bias. 

\citetalias{LoVerde_bias}'s examination of the Lagrangian bias evaluated at a halo's redshift of collapse shows $\frac{d \dcrit}{d \dcbl}$ is nearly independent of scale and halo mass; this feature allows us to write the Lagrangian bias only as a function of redshift and to describe the large-scale amplitude of the halo bias as the galaxy bias of a given sample of galaxies. We then approximate the term derived in \citetalias{LoVerde_bias} to define $\bar{b}(z) \equiv 1+\frac{\partial \ln n(z)}{\partial \dcrit(z)}\frac{d \dcrit(z)}{d \dcbl(z)} $ and express the linear bias relation as
\begin{equation}
\label{eqn::z-seperated-dg}
    \dg (k,\zobs) = \bar{b}(z) \dcbl (k,\zobs). 
\end{equation}
In this form, the halo mass dependence is approximated as an averaged galaxy bias for a given galaxy sample. Further, the expected scale-independent behavior of this form is retrieved in the large-scale limit and for massless-neutrino cosmological models. 

\begin{figure}
    \centering
    \includegraphics[width =0.7\textwidth]{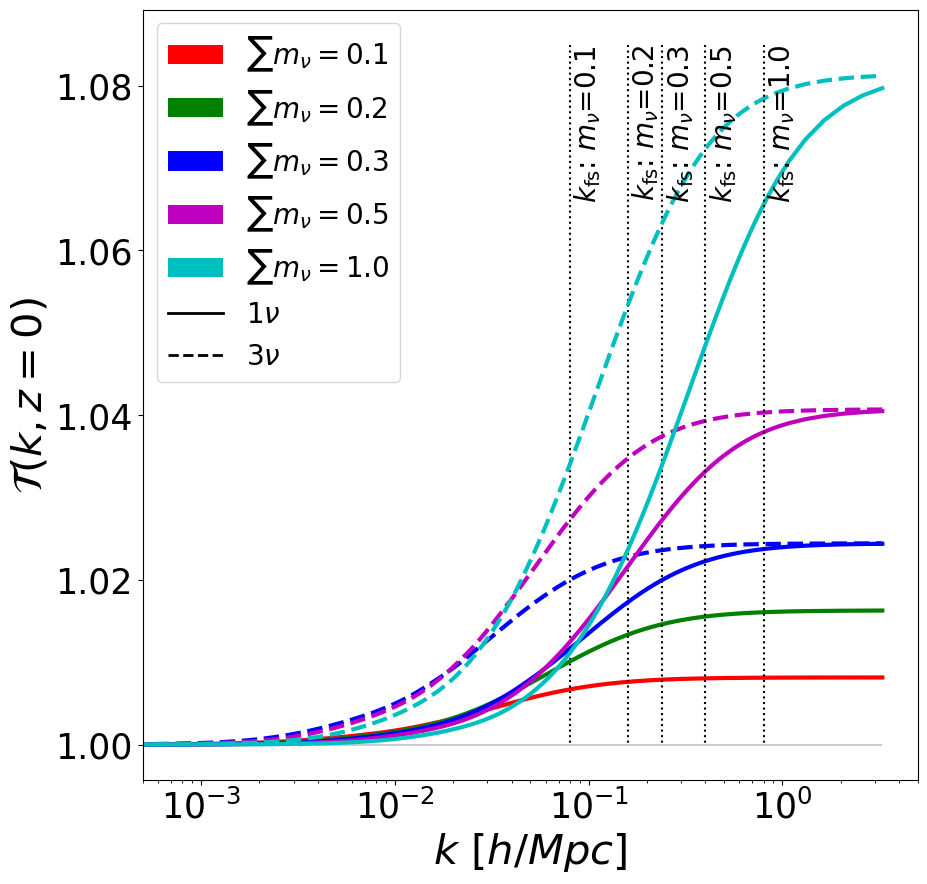}
    \caption{The scale-dependent contribution of the linear galaxy bias as in eq. \eqref{eqn::approx-bias}. 1 massive ($1\n$) and 3 degenerate-mass ($3\n$) neutrino mass models are indicated by solid and dashed lines, respectively, and models evaluated at the same total neutrino mass are color-coordinated. The dotted vertical lines correspond to the free-streaming scale of a single massive neutrino in implicit units of eV. Models with the same total neutrino mass result in an identical increase to the linear galaxy bias at sufficiently small scales. However, models with a greater number of massive species increase the amplitude of the linear galaxy bias at larger physical scales. }
    \label{fig:bias-mu}
\end{figure}

To relate $\dg$ to $\dm$, we adopt \citetalias{LoVerde_bias}'s definition of an apparent linear galaxy bias
\begin{equation}
\label{eqn::lin-matter-gal-bias}
    b(k,z) \equiv \frac{\plin_\gm(k,\zobs)}{\plin_\m(k,\zobs)} = \bar{b}(z) \frac{\plin_{\cb, \m}(k,\zobs)}{\plin_\m(k,\zobs)},
\end{equation}
with $\plin_{X,Y}(k,z)$ the linear power spectrum of fields $\delta_\mathrm{X}$ and $\delta_\mathrm{Y}$ at a given scale, $k$, and redshift, $\zobs$. We denote auto-power spectra as $\plin_\mathrm{X}\equiv \plin_\mathrm{X,X}$ for compactness.

To simplify this expression further, we decompose $\dm$ into neutrino and $\cb$ components,
\begin{equation}
\label{eqn::density-fluctuation}
   \dm(k,\zobs) = \fnu \dnu(k,\zobs) + \fcb \dcb(k,\zobs), 
\end{equation}
where $f_\mathrm{X}$ is the fractional contribution of component $\mathrm{X}$ in the total matter energy density (i.e. $f_\mathrm{X} = \Omega_\mathrm{X}/\Omega_\m$) and the subscript $\n$ corresponds to the contribution of massive neutrinos. We recast the form of eq. \eqref{eqn::lin-matter-gal-bias} using the decomposition in eq. \eqref{eqn::density-fluctuation} to further illustrate the neutrino auto- and cross-power spectra dependencies over all scales as
\begin{equation}
\label{eqn::expanded-bias}
    b(k,\zobs) = \bar{b}(\zobs)\Bigg[\frac{1 + \fcb\frac{\plin_\cb(k,\zobs)}{\plin_\m(k,\zobs)}}{1+\fcb} + \frac{\fnu^2}{1+\fcb} \bigg(\frac{\plin_\cbnu(k,\zobs) - \plin_\n(k,\zobs)}{\plin_\m(k,\zobs)}\bigg)\Bigg].
\end{equation}

Massive neutrino density fluctuations are coherent with those of $\cb$ below the free-streaming scale, $k_\mathrm{fs}$, which allows us to simplify this expression. The free-streaming scale of a single massive neutrino depends on the neutrino's mass and temperature at a given redshift and cosmology and can be written (e.g. as in \citep{Lesgourgues_2012}) as
 \begin{equation}
 \label{eqn::kfs}
     k_{\mathrm{fs}}(z) = 0.8\frac{\sqrt{\Omega_\Lambda + \Omega_\m(1+z)^3}}{(1+z)^2}\bigg(\frac{m_\nu}{1 \mathrm{eV}} \bigg) h \, \mathrm{Mpc}^{-1}.
 \end{equation}
$k_\mathrm{fs}(z)$ decreases as $z \rightarrow 0$ in accordance with an expanding universe with wave numbers $k \gtrapprox k_\mathrm{fs}(z)$ denoting the region where neutrinos propagate freely without interaction. 

Auto- and cross-power spectra with neutrinos are approximately equal where $k \ll k_{\mathrm{fs}}(z)$ (where neutrino density fluctuations are coherent with those of $\cb$) and where $k \gg k_{\mathrm{fs}}(z)$ (where the random motion of neutrinos is uncorrelated with both $\cb$ and neutrino density fluctuations). In these limits, the second term in the square parenthesis of eq. \eqref{eqn::expanded-bias} provides a vanishingly small contribution to the galaxy bias. Outside of these limits, this term is reduced by a factor of $\fnu^2/(1+ \fcb)$. Including this term at the adopted fiducial cosmology (table \ref{tab::fiducial}) impacts the magnitude of the galaxy bias by less than 0.1\%. Our next approximation ignores neutrino clustering contributions to arrive at our final simplified linear galaxy bias model:
\begin{equation}
\label{eqn::approx-bias}
\begin{aligned}
  b(k,\zobs) & \approx \bar{b}(\zobs)  \frac{1+\fcb\frac{\plin_\cb(k,\zobs)}{\plin_\m(k,\zobs)}}{1+\fcb} \\ & = \bar{b}(\zobs) \nisdb (k, \zobs).
\end{aligned}
\end{equation}

In the following analysis, we consider two neutrino mass models: 1 massive neutrino and 3 degenerate-mass neutrinos.\footnote{Additional massless neutrinos are modeled to maintain 3 standard model neutrino species where relevant in our cosmological calculations (i.e. the 1 massive neutrino scenario).} We denote the number of modeled massive neutrinos and the total neutrino mass of the model as 

\begin{align*}
   \mathrm{[\#\,of\,massive\,neutrinos]}\nu \sum m_\n =  \mathrm{[total\,mass\,in\,eV]}.
\end{align*}

Figure \ref{fig:bias-mu} shows the approximation of the NISDB at $z=0$ in a flat $\Lambda$CDM cosmology with massive neutrinos, where we maintain a constant $\Omega_{\m}$ when varying the neutrino mass model. We also plot the associated free-streaming scales of massive neutrinos as in eq. \eqref{eqn::kfs}. We show the galaxy bias begins to increase at scales smaller than the free-streaming scale of the \textit{individual} neutrino masses, but the \textit{total} mass dictates the amplitude of the galaxy bias at small physical scales. The notable impact of the neutrino mass model (1 massive vs. 3 degenerate-mass neutrinos) on the galaxy bias needs to be included in LSS analyses aiming to constrain the neutrino mass hierarchy.

\begin{figure*}
\centering
    \includegraphics[width =0.95\textwidth]{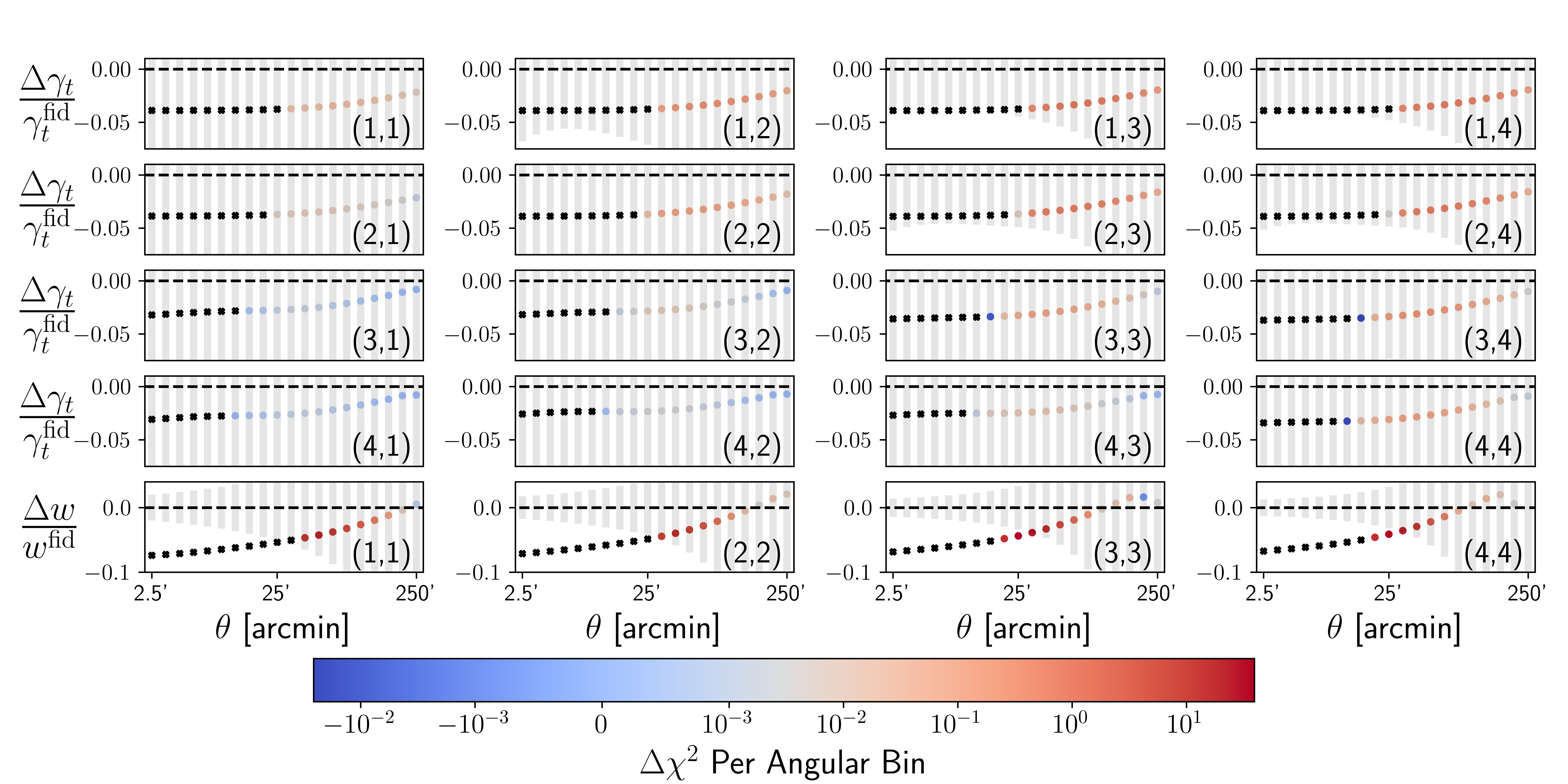}
    \caption{Ratio of noiseless synthetic datavectors in a DESY3 analysis at the adopted fiducial cosmology when utilizing our NISDB model compared to a constant linear galaxy bias in each lens tomographic bin. Excess clustering in the constant linear galaxy bias model corresponds to a positive value of the residual. Scales cut from the analysis due to uncertainties in modeling non-linear galaxy biasing are depicted in black. The color of the points indicates the $\Delta \chi^2$ of each angular bin with values indicated in the color bar below. The total $\Delta \chi^2$ for GGL ($\gamma_t$) and galaxy clustering ($w$) statistics are 2.48 and 11.87, respectively.}
    \label{fig:bias-mu-dataspace}
\end{figure*}

\begin{figure}
\centering
    \includegraphics[width =0.7\textwidth]{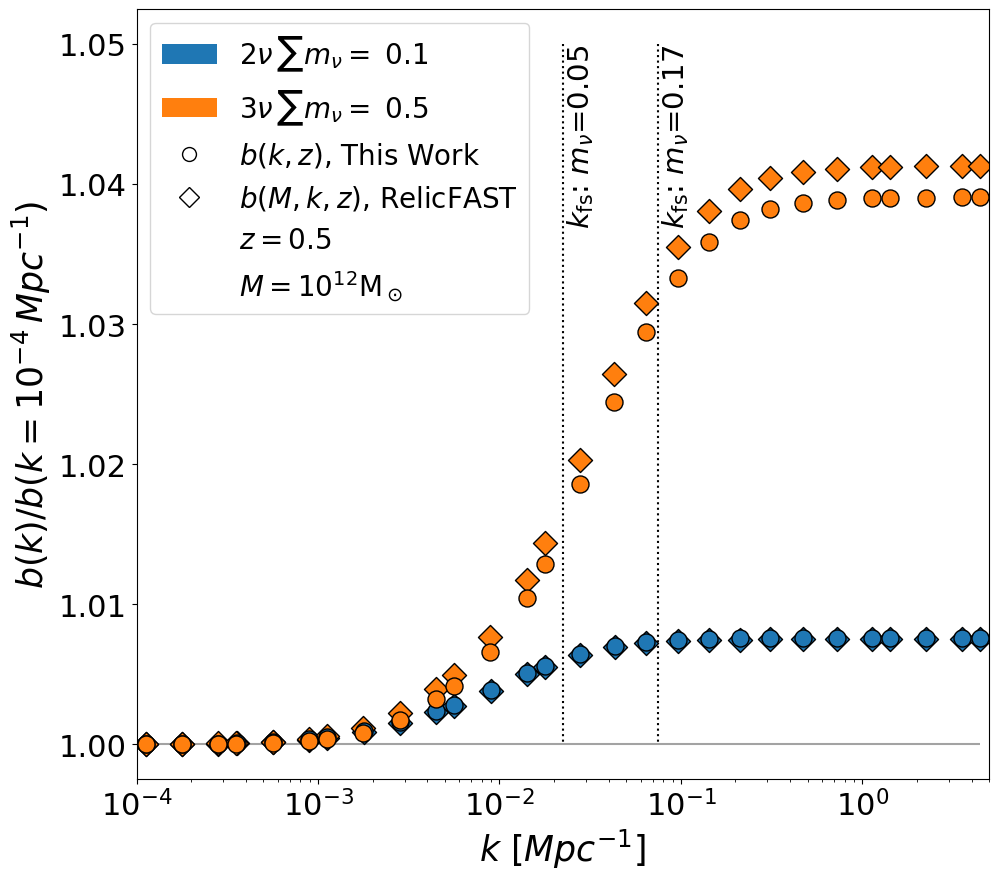}
    \caption{Comparison of the approximate form of the NISDB, eq. \eqref{eqn::approx-bias}, to the linear Eulerian bias using the \texttt{RelicFAST} code for two neutrino mass models, normalized by its large-scale amplitude. The Eulerian bias calculated through \texttt{RelicFAST} utilizes the Sheth-Tormen HMF \citep{Sheth_1999} at $\zcol=0.5$ with a halo mass of $10^{12} M_\odot$. The dotted vertical lines correspond to the free-streaming scale of a single massive neutrino in implicit units of eV. Calculations of the galaxy bias using eq. \eqref{eqn::approx-bias} (circle points) and the linear Eulerian bias from \texttt{RelicFAST} (diamond points) are color-coordinated for like neutrino mass models.} 
    \label{fig:lagrangian}
\end{figure}

Figure \ref{fig:bias-mu-dataspace} illustrates the impact of neutrino modeling on 2pt measurements in configuration space. Specifically, we show the galaxy-galaxy lensing (GGL) and galaxy clustering residuals between synthetic DESY3 datavectors modeled with/without the $\nisdb$ term of eq. \eqref{eqn::approx-bias} at the same cosmology. As $\nisdb(k,z)$ increases with $k$, we expect the largest deviations from the fiducial data to be primarily in galaxy clustering with modifications to GGL statistics. Deviations between the two galaxy clustering datavectors increase towards smaller angular separation, meeting or surpassing the DESY3 error bars. Deviations increase in higher-redshift bins as massive neutrinos free-stream at larger physical scales relative to lower-redshift bins. 

We validate the accuracy of our approximation by comparing it to the halo bias calculated by the spherical-collapse code, \texttt{RelicFAST}, at the adopted fiducial cosmology. \texttt{RelicFAST} includes neutrino self-clustering and Lagrangian bias contributions to calculate the full shape of the halo bias in the presence of massive neutrinos at a given cosmology, halo mass, and HMF. We compare the galaxy bias as in eq. \eqref{eqn::approx-bias} to \texttt{RelicFAST} calculations of the halo bias at the approximate median halo mass for the DESY3 \texttt{MAGLIM} galaxy sample \citep{Porredon_2021}. We use an upper-bound redshift of the collapse of $z = 0.5$, found through the Press-Schechter formalism \citep{1974ApJ...187..425P} at the fiducial DESY3 cosmology with massless neutrinos, as input to the \texttt{RelicFAST} calculation. 

As shown in figure \ref{fig:lagrangian}, the difference between the \texttt{RelicFAST} computation and our approximation is less than 2\% when assuming a minimal-mass normal hierarchy. Our approximation agrees less with \texttt{RelicFAST} at larger $\sum m_\n$. For the $3 \n \sum m_\n = 0.5$ case, the halo bias begins to increase at the typical scales of halo collapse and is $\sim$ 20\% larger than our approximation at small physical scales. When analyzing deviations between our approximation and the full expression through \texttt{RelicFAST}, it is important to note that non-linear evolution and astrophysical effects (baryonic feedback, non-linear galaxy clustering, etc.) dominate at these small physical scales. Such scales with high model uncertainty are not well-described by linear perturbation theory and are typically excluded from analyses through scale-cuts. We find eq. \eqref{eqn::approx-bias} accurately captures the dominant terms of the full NISDB for the most commonly observed halo mass of the \texttt{MAGLIM} sample across the relevant scales used in DESY3 cosmological analyses. The median halo mass is expected to be lower for LSST-like sensitivity, reducing discrepancies between the approximate and \texttt{RelicFAST} calculations of the NISDB.

\section{Analysis and Modeling Choices}
\label{sec::pipeline}
 
\begin{table}
\label{tab::fiducial}
\begin{center}
\begin{tabular}{|l l|}
\hline
Parameter & Fiducial  \\ \hline
\multicolumn{2}{|c|}{\textbf{Cosmology}} \\ 
$\Omega_{\rm\m} $ & $0.3$\\
$10^{-9}A_{\rm{s}}$ & $2.19$\\
 
$\Omega_{\rm{b}}$ & $0.048$\\

$n_{\rm{s}}$ & $0.97$\\

$h$ & $0.69$\\

$10^{-4} \Omega_\n h^2$& $53.7$\\
$\n$ & $3$\\

\hline
\multicolumn{2}{|c|}{\textbf{Intrinsic Alignment}} \\ 

 $a_1$ & $0.7$\\
 $\eta_1$ & $-1.7$\\

\hline
 \multicolumn{2}{|c|}{\texttt{MAGLIM}\ \textbf{Galaxy Bias}} \\  
$\bar{b}^{1\cdots 4}$  & $1.7,1.8,1.8,1.9$\\
 \hline

\multicolumn{2}{|c|}{LSST\ \textbf{Galaxy Bias}} \\  
$\bar{b}^{1\cdots 5}$  & $1.7,1.7,1.7,2.0,2.0$\\
 \hline
\end{tabular}
\end{center}
\caption{Fiducial cosmological and nuisance parameters utilized in this analysis. $\n$ denotes the number of massive neutrinos in the adopted fiducial model (i.e. 3 degenerate-mass neutrinos). Parameters with unique fiducial values in each bin are listed from lowest to highest in redshift.}
\end{table}

By measuring both galaxy positions and galaxy shapes as a function of redshift, photometric redshift galaxy surveys like DES and LSST combine galaxy clustering and cosmic shear 2pt statistics with their GGL cross-correlations to place constraints on cosmological parameters, known as the 3x2pt analysis. This section details our modeling pipeline, where we assume a flat $\Lambda$CDM universe with 3 degenerate-mass neutrinos as the adopted fiducial cosmology (see table \ref{tab::fiducial}).

We denote quantities associated with the galaxy clustering statistics/galaxy lens sample with a subscript $\g$, while those concerning galaxy shear statistics/galaxy source sample have the subscript $\kap$. A capital letter (e.g. $\mathrm{A}$ or $\mathrm{B}$) will correspond to either $\g$ or $\kap$ when eqs.  are written with generality.  Superscript Latin indices (e.g. $i$ and $j$) denote the tomographic bin number of a galaxy sample. $n^i_\mathrm{A}(z)$ is defined to be the observed number density of galaxies in tomographic bin $i$ for a given lens/source galaxy sample, $\mathrm{A}$. z-averaged quantities of tomographic bin $i$ are barred, where we make particular use of the average redshift, $\bar{z}^i$, and average number density of galaxies, $\bar{n}_{\mathrm{A}}^i$, of a galaxy sample. To shorten notation, we take the redshift, $z$, to be an implicit function of comoving distance, $\chi$, as $z \equiv z(\chi)$. 
For our DESY3 analysis, we utilize the DESY3 source sample \citep{Gatti_2021} and \texttt{MAGLIM} lens sample \citep{Porredon_2021}, each containing four tomographic bins as in the fiducial DESY3 analysis \citep{y3-cosmology}. Our LSSTY1 analysis follows the $n(z)$ assumptions of \cite{lsst-srd} to generate source and lens samples each divided into five tomographic bins, as expected for LSSTY1 \citep{Fang_2020}. In addition to binned auto-correlations of galaxy clustering, 7 GGL bins are analyzed in the LSSTY1 analysis in accordance with \cite{lsst-srd} and all 16 lens-source combinations are considered in the DESY3 analysis \citep{y3-modeling}. 

\subsection{Computing 2pt Correlation Functions (2PCFs)}
We focus on statistics involving galaxy clustering in this section and refer the reader to Sections 2--4 of the DESY3 modeling paper \citep{y3-modeling} for further details on the weak-lensing model of the 3x2pt datavector. 

The angular 2D correlation function of two tracer fields for a Fourier mode, $l$, is calculated as a line-of-sight projection of the 3D galaxy/matter density contrast weighted by tracer projection kernels, $q_\mathrm{A}(k,\chi)$. Using the Limber approximation, this is written as 
\begin{equation}
\label{eqn::limber}
    C_{\mathrm{AB}}^{ij}(l) = \!\! \int\!\! d\chi \frac{q_{\mathrm{A}}^i\!\!\left(\!\frac{l+1/2}{\chi},\chi\right)q_{\mathrm{B}}^j\left(\frac{l+1/2}{\chi},\chi\right)}{\chi^2}  \pnl_\m\left(\frac{l+1/2}{\chi},z\right), 
\end{equation}

where $P_\m^\mathrm{nl}$ is the total matter non-linear power spectrum computed through \texttt{HALOFIT} \cite{Takahashi_2012}.

The galaxy clustering projection kernel, $q_\g(k,\chi)$, is defined with the approximate NISDB of eq. \eqref{eqn::approx-bias} in terms of comoving distance for a given redshift bin, $i$, as 
\begin{equation}
\label{eqn::lensing-kernel}
    q_\g^i(k,\chi) = \bar{b}^i \nisdb\left(k, z\right) \frac{n_{\g}^i(z) }{\bar{n}_{\g}^i}\frac{dz}{d\chi} + C_\mathrm{M}^i W_\mathrm{\g}^i(\chi)\,,
\end{equation}
with $\bar{b}^i$ the large-scale linear galaxy bias of tomographic bin $i$, where $\bar{b}^i$ is written as $\bar{b}(\zobs)$ in eq. \eqref{eqn::approx-bias}. The second term is the magnification contribution to the projected density contrast, with $C_\mathrm{M}^i$ the lens magnification bias and $W_\mathrm{A}^i(\chi)$ the lens efficiency of galaxy sample $\mathrm{A}$,
\begin{equation}
    W_\mathrm{A}^i(\chi) = \frac{3\Omega_\m H_0^2}{2c^2}\int_\chi^{\infty} d\chi' \frac{n_\mathrm{A}^i(z')}{\bar{n_\mathrm{A}}^i}\frac{\chi}{a(\chi)}\frac{\chi'-\chi}{\chi'}
\end{equation}
with the Hubble Constant $H_0$, speed of light $c$, fractional energy density of matter $\Omega_\m$, and scale-factor $a(\chi)$.

We adopt the NLA model (detailed in section \ref{sec::systematics}), which allows us to write the Intrinsic Alignment (IA) contribution, $q_\kapI(\chi)$, as an additive factor of the shear projection kernel, $q_\kap^i(\chi)$. The shear projection kernel is then defined as
\begin{equation}
\label{eqn::shear-kernel}
    q_{\kap}^i(\chi) = W_\mathrm{\kap}^i(\chi) + q_\kapI(\chi).
\end{equation}

At large physical scales, the Limber Approximation used in eq. \eqref{eqn::limber} does not hold. We then use the framework developed by \cite{Fang_2020} to efficiently compute the angular clustering power spectrum without the Limber approximation. The speed and accuracy of the non-Limber calculation rely on the integrals of correlation functions to be separable in $\chi$ and $k$ at large scales. The form of the NISDB in eq. \eqref{eqn::approx-bias} can be made separable in $\chi$ and $k$ by evaluating $\nisdb(k,z)$ at the mean redshift of each tomographic bin, which is a good approximation for the sufficiently narrow redshift bins of the LSSTY1 and DESY3 lens samples. We separate the linear growth factor, $G(z)$, from the total matter linear power spectrum to write


\begin{align}
    \nonumber C_{\g\g}^{ii}(l) = & \frac{2}{2l+1}\int d\chi \frac{q_\g^i\!\!\left(\!\frac{l+1/2}{\chi},\chi\right)q_\g^i\!\left(\frac{l+1/2}{\chi},\chi\right)}{\chi^2}\left[\pnl_\m \left(\frac{l+1/2}{\chi},z \right) - \plin_\m \left(\frac{l+1/2}{\chi},z\right)\right] \\ + & \frac{2}{\pi}\int d\chi_1 G(z_1)\int d\chi_2 G(z_2) \int_0^\infty \frac{dk}{k}k^3 \plin_\m(k,0)q_\g^i(k, \chi_1) q_\g^i(k, \chi_2)j_l(k\chi_1)j_l(k\chi_2),
\label{eqn::fftlog}
\end{align}
where $j_l(x)$ is the spherical Bessel function.

DESY3 analyzes configuration-space correlation functions for a given angular separation in the sky, $\theta$, in galaxy clustering, $w^i(\theta)$, and GGL, $\gamma_{\mathrm{t}}^{ij}(\theta)$. For these analyses, $C_{\mathrm{A}\mathrm{B}}^{ij}(l)$'s are converted into configuration-space angular correlation functions as
\begin{equation}
\begin{aligned}
\label{eqn::ang-2pt}
    w^{i}(\theta) &= \sum_l \frac{2l+1}{4\pi} \mathcal{P}_l\left(\cos(\theta)\right)\,C_{\g\g}^{ii}(l)\,\\
     \gamma_{\mathrm{t}}^{ij}(\theta) &=\sum_l \frac{2l+1}{4\pi l(l+1)} \mathcal{P}_l^2\left(\cos(\theta)\right) C_{\g\kap}^{ij}(l) \,
\end{aligned}
\end{equation}
with $\mathcal{P}_l(x)$ the Legendre polynomial and $\mathcal{P}_l^2(x)$ the associated Legendre polynomial of order $l$. 

\subsection{Galaxy Clustering and Lensing Systematics}
\label{sec::systematics}
Several systematic effects need to be included to connect the theoretical model derived in the previous subsection to observations. The implementation of these systematic parameters is summarized in this subsection. Prior ranges and fiducial values of sampled parameters are listed in table \ref{tab::priors}.

\begin{table}
\label{tab::priors}
\begin{center}
\begin{tabular}{|l r|}
\hline
Parameter & Prior  \\ \hline
\multicolumn{2}{|c|}{\textbf{Cosmology}} \\ 
$\Omega_{\rm\m} $ & $\mathcal{U}[0.1, 0.9]$\\
$10^{-9}A_{\rm{s}}$ & $\mathcal{U}[0.5, 5.0]$\\
 
$\Omega_{\rm{b}}$ & $\mathcal{U}[0.03, 0.07]$\\

$n_{\rm{s}}$ & $\mathcal{U}[0.87, 1.06]$\\

$h$ & $\mathcal{U}[0.55, 0.91]$\\

$10^{-4} \Omega_\n h^2$& $\mathcal{U}[6.0, 107.4]$\\  

\hline
\multicolumn{2}{|c|}{\textbf{Intrinsic Alignment}} \\ 

 $a_1$ & $\mathcal{U}[-5.0, 5.0]$\\
 $\eta_1$ & $\mathcal{U}[-5.0, 5.0]$\\

\hline
\hline
 \multicolumn{2}{|c|}{\texttt{MAGLIM}\ \textbf{Galaxy Bias}} \\  
$\bar{b}^{1\cdots 4}$  & $\mathcal{U}[0.8, 3.0]$\\

\hline
 \multicolumn{2}{|c|}{\texttt{MAGLIM}\ \textbf{Lens Magnification Bias}} \\  
 $C_\mathrm{M}^{1\cdots 4}$ & $0.43,\, 0.30,\, 1.75,\, 1.94$ \\

\hline
  \multicolumn{2}{|c|}{\texttt{MAGLIM}\ \textbf{Lens Mean Redshift Uncertainty}} \\  
$\Delta_l^{1 \cdots 4}$  & $\mathcal{G}[-0.009, 0.007],\, \mathcal{G}[-0.035, 0.011]$, $\mathcal{G}[-0.005, 0.006],\, \mathcal{G}[-0.007, 0.006]$\\
\hline
  \multicolumn{2}{|c|}{DESY3 \textbf{Source Mean Redshift Uncertainty}} \\  
$\Delta_s^{1\cdots 4}$  & $\mathcal{G}[0.0, 0.018],\, \mathcal{G}[0.0, 0.013]$, $\mathcal{G}[0.0, 0.006],\, \mathcal{G}[0.0, 0.013]$\\
\hline
  \multicolumn{2}{|c|}{DESY3 \textbf{Shear Calibration Bias}} \\  
$m^{1\cdots 4}$  & $\mathcal{G}[-0.006, 0.008],\, \mathcal{G}[-0.01, 0.013]$, $\mathcal{G}[-0.026, 0.009], \, \mathcal{G}[-0.032, 0.012]$\\

\hline
  \multicolumn{2}{|c|}{\texttt{MAGLIM}\ \textbf{Lens Redshift Width Uncertainty}} \\  
$s^{1\cdots 4}$  & $\mathcal{G}[0.975, 0.06],\, \mathcal{G}[1.306, 0.09]$, $\mathcal{G}[0.870, 0.05],\, \mathcal{G}[0.918, 0.05]$\\

\hline
\hline
 \multicolumn{2}{|c|}{LSST\ \textbf{Galaxy Bias}} \\  
$\bar{b}^{1\cdots 5}$  & $\mathcal{U}[0.8, 3.0]$\\

\hline
 \multicolumn{2}{|c|}{LSST\ \textbf{Lens Magnification Bias}} \\  
$C_\mathrm{M}^{1\cdots 5}$ & $-0.19,\, -0.63,\, -0.69,\, 1.18, \, 1.88$\\

\hline
  \multicolumn{2}{|c|}{LSST\ \textbf{Lens Mean Redshift Uncertainty}} \\  
$\Delta_l^{1\cdots 5}$  & $\mathcal{G}[0.0, 0.005]$\\
\hline
  \multicolumn{2}{|c|}{LSST\ \textbf{Source Mean Redshift Uncertainty}} \\  
$\Delta_s^{1\cdots 5}$  & $\mathcal{G}[0.0, 0.002]$\\
\hline
  \multicolumn{2}{|c|}{LSST\ \textbf{Shear Calibration Bias}} \\  
$m^{1\cdots 5}$  & $\mathcal{G}[0.0, 0.013]$\\

\hline
  \multicolumn{2}{|c|}{LSST\ \textbf{Lens Redshift Width Uncertainty}} \\  
$s^{1\cdots 5}$  & $\mathcal{G}[1.0, 0.1]$\\
 \hline
\end{tabular}
\end{center}
\caption{Table of parameter priors utilized in this analysis. Gaussian priors are denoted as $\mathcal{G}$[mean, uncertainty] and uniform priors as $\mathcal{U}$[lower-bound, upper-bound], while lists of values denote fixed parameters. Parameters with distinct priors for each redshift bin are listed from lowest to highest in redshift, while parameters with identical priors in each bin are written once for conciseness. We additionally marginalize over a point-mass for each lens bin in DESY3 GGL statistics that is not listed in the table below, where the fiducial value and prior for each bin is taken to be $0$ and $\mathcal{U}[-100,100]$, respectively.}
\end{table}

\textbf{Galaxy Bias}
We parameterize the per-bin large-scale galaxy bias amplitude as $\bar{b}^i$. Our analysis also considers a linear galaxy bias model where galaxies trace $\dcb$. Modulating the galaxy bias by eq. \eqref{eqn::approx-bias} only traces the $\dcb$ field in the linear regime. Using galaxies as tracers of the underlying non-linear $\dm$ field with the $\nisdb(k,z)$ correction differs at small scales compared to tracing the non-linear $\dcb$ field. To properly model the linear galaxy bias to $\cb$ density fluctuations, we take $\nisdb(k,z) = 1$ and $q_\gD^i(\chi) \rightarrow q_\gD^i(\chi)\sqrt{\pnl_{\cb}(k,z)/\pnl_{\m}(k,z)}$ or $q_\gD^i(\chi) \rightarrow q_\gD^i(\chi)\sqrt{\plin_{\cb}(k,z)/\plin_{\m}(k,z)}$ where appropriate in eqs.  \eqref{eqn::limber} and \eqref{eqn::fftlog}.

\textbf{Galaxy Redshift Distribution} Uncertainties in the shape of the redshift distribution of galaxies are parameterized by an additive shift to the mean redshift of the distribution and a multiplicative stretch that modulates the width of the redshift distribution. We follow the DESY3 parameterization \citep{cawthon2020dark} where a given redshift distribution of tomographic bin $i$, $n_\mathrm{A}^i(z)$, may be shifted by $\Delta^i$ and stretched by $s^i$ as 
\begin{equation}
    n_\mathrm{A}^i(z) \rightarrow \frac{1}{s^i}n_\mathrm{A}^i\!\left( \frac{z - \bar{z}^i - \Delta^i}{s^i} + \bar{z}^i\!\right).
\end{equation}
Shifts in the mean redshift distributions are implemented for both the source and lens galaxy samples in each tomographic bin as $\Delta^i_s$ and $\Delta^i_l$, respectively. We sample the redshift distribution stretch only for lens sample bins and fix $s^i=1$ for source sample bins.

\textbf{Lens Magnification and Shear Calibration}
To account for the magnification of lens sample galaxies, a lens magnification bias parameter, $C^i_\mathrm{M}$, is modeled for each lens tomographic bin and is kept fixed in the analysis (as validated in \citep{desy3-mag}). We also fix this value for the LSSTY1 analysis for computational simplicity. 

We parameterize a multiplicative shear calibration for the source galaxy sample, where the shear projection kernel of eq. \eqref{eqn::shear-kernel} transforms as 
\begin{equation}
    q^i_\kap(\chi) \rightarrow (1+m^i)q^i_\kap(\chi).
\end{equation}
The magnitude of the shear calibration bias parameter, $m^i$, is varied for each redshift bin of the source sample.

\textbf{Intrinsic Alignments}
The intrinsic alignment of galaxies is modeled as an additive systematic contaminating the observed weak lensing signal of source galaxies. We use the NLA model \citep{Catelan_2001, PhysRevD.70.063526, Bridle_2007}, parameterizing an amplitude ($a_1$) with a power-law scaling in redshift ($\eta_1$), to characterize preferential alignment of galaxies to their local environment. First introduced in eq. \eqref{eqn::shear-kernel}, we now write the form of $q^i_\kapI(\chi)$ as
\begin{equation}
    q^i_\kapI(\chi) = -a_1 \!\left( \frac{1+z}{1+z_0}\!\right)^{\eta_1} \frac{\bar{C_1}\rho_\mathrm{crit}\Omega_\m}{G(z)}\frac{n^i_\kap(z)}{\bar{n}^i_\kap}.
\end{equation}
This IA model differs from the DESY3 fiducial analysis which included higher-order tidal effects through the TATT model \citep{Blazek_2015, Blazek_2019}. Analysis of the Y3 data, however, showed little IA model preference, so we use the NLA model for simplicity \citep{Secco_2022}.  

\textbf{Non-Local Shear}
For $\gamma_{\mathrm{t}}^{ij}(\theta)$ statistics, the projected mass around lens galaxies shears source galaxy shapes on scales smaller than $\theta$. We parameterize this contribution using the point-mass marginalization model \citep{MacCrann_2019} with uninformative flat priors of $[-100,100]$ to account for non-local mass contributions in the projected mass and take the fiducial point-mass value to be 0 for configuration-space DESY3 simulated analyses.

\subsection{Likelihood Analysis}
\label{sec::like}
We generate theory datavectors of a given model to sample all free parameters in a model and assume a Gaussian likelihood 
\begin{equation}
\label{eqn::like}
\mathcal{L} \propto \exp \biggl(-\frac{1}{2} \left[ \left(\D -\M\right)^\nisdb \, \matC^{-1} \, \left(\D-\M\right) \right]  \biggr).
\end{equation}
The likelihood function, $\mathcal{L}$, depends on the observed simulated datavector, $\D$, the model datavector, $\M$, and the pre-computed covariance matrix, $\matC$. The term in brackets is the commonly-defined $\Delta \chi^2$. The DESY3/LSSTY1 covariance matrices are computed at the fiducial cosmology with one massive neutrino at minimal mass following \cite{Fang__2020} in configuration\footnote{https://github.com/CosmoLike/CosmoCov}/Fourier\footnote{https://github.com/CosmoLike/CosmoCov\_Fourier} space, respectfully. When computing the LSSTY1 Fourier-space covariance matrix, we marginalize over nuisance parameters with well-constrained Gaussian priors (i.e. shear calibration, lens/source photo-z uncertainty, and lens stretch) analytically and incorporate their contributions at the covariance level. Table \ref{tab::fiducial} provides the fiducial cosmology and table \ref{tab::priors} the priors of sampled parameters. We take the fiducial value of parameters with Gaussian priors to be the prior's expectation value.

Datavectors and likelihood evaluations are computed using \texttt{CosmoLike} \cite{Krause_2017} with \texttt{CLASS} \cite{CLASS} and we sample parameter space with \texttt{MULTINEST}\footnote{https://github.com/JohannesBuchner/MultiNest} nested sampling. We follow the guidelines for setting \texttt{MULTINEST} hyperparameters as examined in \cite{Lemos_2022, Miranda_2021} to provide sufficient sampling of all parameters. 

The fiducial DESY3 scale-cuts for galaxy clustering were designed to exclude scales affected by unmodeled systematic effects, including non-linear galaxy biasing, from the cosmological analysis. We follow the same procedures as DESY3 \citep{y3-modeling} to test the validity of the fiducial scale-cuts when using the approximate NISDB. We verify that at these scale cuts the $\Delta \chi^2<1$ without refitting for two data vectors with and without non-linear galaxy biasing terms (fixed to their co-evolution values, as in \citep{PhysRevD.90.123522}) and with approximate NISDB at $3 \nu \sum m_\n = 1.0$ (corresponding to the upper-limit of the $\Omega_\n h^2$ prior range). This indicates that the fiducial scale-cuts omit non-linear galaxy biasing sufficiently well for our DESY3 synthetic likelihood analysis. For LSSTY1 analyses, we adopt the scale-cuts from \cite{lsst-srd, Fang_2020}. 

\begin{table*}
\centering
\begin{tabular}{|p{40mm}|P{40mm}|}\hline 
\backslashbox[44mm]{Analysis \\ Model}{Fiducial \\ Model} & $3\n, \sum m_\n=0.5$  \newline \baseline\\
\hline
$3\n$ \newline \baseline & baseline \\\hline
$3\n$ \newline  \bmat & \textbf{(1)} bias model \\\hline
$1\n$ \newline  \bmat & \textbf{(2)} bias model \& $m_\n$ mass model \\\hline
$1\n$ \newline  \baseline & \textbf{(3)} $m_\n$ mass model \\\hline
$3\n$ \newline  \bcb & \textbf{(4)} underlying field galaxies trace \\\hline
$1\n$ \newline  \bcb & \textbf{(5)} underlying field galaxies trace \&  $m_\n$ mass model\\
\hline
\end{tabular}
\caption{Models used to generate the input datavector, $\D$, and model datavector, $\M$, in our likelihood analysis. The input datavector is generated at the fiducial cosmology in table \ref{tab::fiducial} with the approximate NISDB of eq. \eqref{eqn::approx-bias}. Cosmological parameters are inferred assuming the neutrino mass and linear galaxy bias models of the top row. Differences between the fiducial model and analysis model are written in the bottom row where bold numbers are used to reference inferences of the analysis model in our discussion.}
\label{tab::neutrino-models}
\end{table*}

\section{Galaxy Bias Model Comparison and Discussion}
\label{sec::results}
We design a number of numerical experiments to quantify the impact of galaxy bias and neutrino mass models on cosmological parameter estimation when data contains an appreciable neutrino-induced scale-dependent bias. The input fiducial synthetic datavector is calculated with the scale-dependent bias following eq. \eqref{eqn::approx-bias} and neutrino mass model $3 \nu \sum m_\n = 0.5$. We maintain the fiducial synthetic datavector, $\D$, for each model comparison.

Each model comparison uses one linear galaxy bias model and one neutrino mass model to generate model data, $\M$. We consider two neutrino mass models (1 massive or 3 degenerate-mass neutrinos) and three linear galaxy bias models ($\dg = \bar{b}^i \dcb$, $\dg = \bar{b}^i \dm$, and the NISDB approximation), totaling six analysis models (five with model misspecification variations), and infer cosmological parameters through a nested sampling analysis. Table \ref{tab::neutrino-models} summarizes the conducted model misspecification analyses. 
\begin{table*}
    \centering
        \begin{tabular}{|l|l|c|c|c|c|c|c|}
        \hline
        \multicolumn{2}{|l|}{\multirow{2}{*}{Model}} &\multicolumn{6}{c|}{1D Parameter Shifts DESY3}\\
        \cline{3-8}
        \multicolumn{2}{|l|}{\multirow{2}{*}{}}& $\Omega_\m$ & $h_0$ & $\sigma_{8,\cb}$ & $\sigma_8$ & $S_8$ & $\sum m_\n$ \\   
        \hline
 \multirow{3}{*}{$3\n$}& \baseline & +0.03 & +0.03 & $-$0.05 & $-$0.05 & $-$0.07 & +0.08  \\
 \cline{2-8}
 & \bmat & +0.37 & $-$0.01 & $-$0.34 & $-$0.32 & $-$0.10 & +0.11 \\
 \cline{2-8}
& \bcb & +0.24 & +0.01 & $-$0.23 & $-$0.23 & $-$0.14 & +0.22  \\
\hline
\multirow{3}{*}{$1\n$}& \baseline & $-$0.04 & +0.14 & $-$0.08 & +0.13 & +0.22 & $-$0.43 \\
\cline{2-8}
& \bmat & $-$0.09 & +0.13 & $-$0.04 & +0.22 & +0.37 & \textbf{$-$1.07}  \\
\cline{2-8}
& \bcb & $-$0.05 & +0.11 & $-$0.11 & +0.13 & +0.22 & \textbf{$-$0.82} \\
\hline\hline

        \multicolumn{2}{|l|}{\multirow{2}{*}{Model}} &\multicolumn{6}{c|}{1D Parameter Shifts LSSTY1}\\
        \cline{3-8}
        \multicolumn{2}{|l|}{\multirow{2}{*}{}}& $\Omega_\m$ & $h_0$ & $\sigma_{8,\cb}$ & $\sigma_8$ & $S_8$ & $\sum m_\n$ \\   
        \hline
 \multirow{3}{*}{$3\n$}& \baseline & $-$0.02 & +0.05 & +0.01 & +0.01 & $-$0.03 & +0.03 \\
 \cline{2-8}
 & \bmat  & \textbf{+0.61} & \textbf{+0.54} & \textbf{$-$0.98} & \textbf{$-$0.55} & $-$0.07 & +0.15 \\
 \cline{2-8}
& \bcb & \textbf{+0.62} & \textbf{+0.64} & \textbf{$-$0.77} & \textbf{$-$0.70} & $-$0.30 & \textbf{+0.50} \\
\hline
\multirow{3}{*}{$1\n$}& \baseline & $-$0.21 & \textbf{+1.00} & \textbf{$-$1.10} & \textbf{+1.35} & \textbf{+2.27} & $-$0.41 \\
\cline{2-8}
& \bmat & +0.24 & \textbf{+1.50} & \textbf{$-$1.88} & \textbf{+0.70} & \textbf{+2.09} & $-$0.43 \\
\cline{2-8}
& \bcb & +0.11 & \textbf{+1.58} & \textbf{$-$1.72} & \textbf{+0.80} & \textbf{+1.87} & $-$0.09\\
\hline        
        \end{tabular}
    \caption{Projected 1D parameter shifts between the fiducial cosmology and MAP cosmology of a given model in DESY3 and LSSTY1 synthetic analyses, presented in terms of the marginalized 1D parameter uncertainty, $\sigma$. Positive and negative values denote shifts above and below the input cosmology, respectively. Parameter shifts above 0.5$\sigma$ are presented in bold.}
    \label{tab:1dparams}
\end{table*}
\begin{table}
    \centering
    \begin{tabular}{|l|l|c|c|c|c|}
        \hline
        \multicolumn{2}{|l|}{\multirow{2}{*}{Model}} &\multicolumn{2}{c|}{$S_8$-$\Omega_m$ } & \multicolumn{2}{c|}{$\sigma_8$-$\sigma_{8,\cb}$}\\
        \cline{3-6}
        \multicolumn{2}{|l|}{\multirow{2}{*}{}}& DESY3 & LSSTY1 & DESY3  & LSSTY1 \\
        \hline
        \multirow{3}{*}{$3\n$}& \baseline & 0.02& 0.01 & 0.07 & 0.01 \\
        \cline{2-6}
        & \bmat & 0.1 & 0.31 & 0.23& \textbf{0.89} \\
        \cline{2-6}
        & \bcb & 0.03 & 0.31& 0.13 & 0.49 \\
        \hline
        \multirow{3}{*}{$1\n$}& \baseline & 0.02 & \textbf{2.34} & \textbf{2.51} & \textbf{10.08} \\
        \cline{2-6}
        & \bmat & 0.03 & \textbf{1.54} & \textbf{2.85} & \textbf{10.08} \\
        \cline{2-6}
        & \bcb & 0.03 & \textbf{1.48} & \textbf{2.73} & \textbf{8.28} \\
        \hline

    \end{tabular}
    \caption{Projected 2D parameter shifts between the fiducial cosmology and the MAP cosmology of a given model in DESY3 and LSSTY1 synthetic analyses, presented in terms of the marginalized 2D parameter uncertainty, $\sigma$. Parameter shifts above 0.5$\sigma$ are presented in bold.}
    \label{tab:2dparam}
\end{table}

\begin{table*}
    \centering
    \begin{tabular}{|l|l|c|c|c|c|c|c|}
        \hline
        \multicolumn{2}{|l|}{\multirow{2}{*}{Model}} &\multicolumn{3}{c|}{$\Delta \chi^2$ DESY3 } & \multicolumn{3}{c|}{$\Delta \chi^2$ LSSTY1 }\\
        \cline{3-8}
        \multicolumn{2}{|l|}{\multirow{2}{*}{}} & 3x2pt & $\gamma_t$ + $w$ & $w$ & 3x2pt & $\gamma_t$ + $w$ & $w$\\   
        \hline
        \multirow{3}{*}{$3\n$}&\baseline & 0.03 & 0.03 & 0.01 & 0.00 & 0.00 & 0.00 \\
        \cline{2-8}
        &\bmat & 0.15 & 0.11 & 0.05 & 0.19 & \textbf{2.40} & \textbf{1.43} \\
        \cline{2-8}
        &\bcb & 0.25 & 0.24 & 0.18 & 0.05 & \textbf{2.33} & \textbf{1.65} \\
        \hline
        \multirow{3}{*}{$1\n$}& \baseline & 0.29 & 0.29 & 0.24 & \textbf{1.84} & 0.62 & 0.48 \\
        \cline{2-8}
        &\bmat & 0.34 & 0.31 & 0.24 & \textbf{2.70} & \textbf{4.60} & \textbf{3.57} \\
        \cline{2-8}
        &\bcb & 0.32 & 0.30 & 0.25 & \textbf{2.74} & \textbf{3.83} & \textbf{3.19} \\
        \hline

    \end{tabular}
    \caption{$\Delta \chi^2$ statistics as compared to the fiducial data vector, where $\Delta \chi^2$ greater than 1 are presented in bold. Anti-correlations between shear and clustering statistics in the LSSTY1 analysis can result in 2x2pt $\Delta \chi^2$ values to be larger than the 3x2pt case. MAP datavectors in the DESY3 analysis are difficult to find precisely due to the number of free parameters in the model and the length of the datavector. These complexities result in a small $\Delta \chi^2$ between the fiducial datavector and the MAP datavector using the fiducial model, where similar effects propagate to other DESY3 MAP comparisons. These model complexities do not impact the finding of LSSTY1 MAP datavectors. }
    \label{tab:dchisq}
\end{table*}

We conduct cosmological inferences for each model listed in table \ref{tab::neutrino-models} and find the Maximum A-Posteriori (MAP) cosmology for each chain using the Nelder-Mead algorithm \citep{Nelder-Mead} in \texttt{scipy}\footnote{https://scipy.org/}. We compare MAP points to the fiducial cosmology and to derived parameters $\sigma_8$ (the amplitude of $\plin_\m(k,0)$ fluctuations at scales of 8 Mpc/h) and $S_8\equiv \sigma_8 \, \sqrt{\Omega_\m}$. We also report marginalized parameter shifts in $\sigma_{8,\cb}$, defined in the same manner as $\sigma_{8}$ but utilizes $\plin_\cb(k,0)$ to compute the amplitude of density fluctuations at 8 Mpc/h scales. 

We choose our systematic bias criteria such that MAP datavectors with a $\Delta \chi^2 >1$ and 2D parameter shifts in the $S_8$ - $\Omega_m$ plane greater than $0.3\sigma$ are considered to be systematically biased. We provide best-fit projected 1D and 2D cosmological parameter shifts in tables \ref{tab:1dparams} and \ref{tab:2dparam} with datavector $\Delta \chi^2$'s in table \ref{tab:dchisq} for each model and survey considered. 

\begin{figure}
    \centering
    \includegraphics[width =0.7\textwidth]{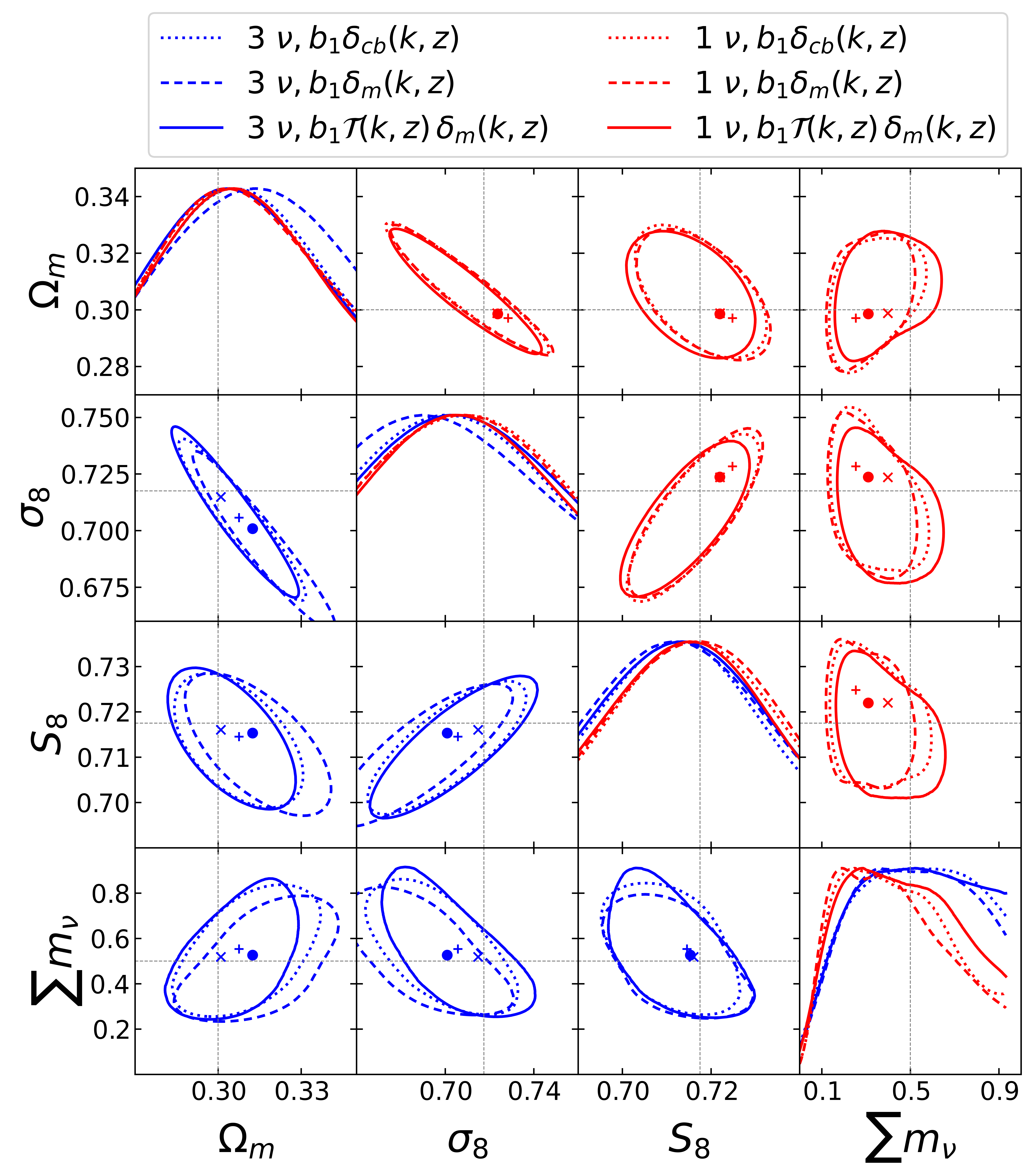}
    \caption{DESY3 simulated $0.3\sigma$ constraints of the marginalized 2D parameter contours for the neutrino mass and galaxy bias models listed in the given legend. Best-fit MAP values are depicted as cross, plus, and circle points for the NISDB, the constant linear galaxy bias model tracing $\dm$, and the constant linear galaxy bias model tracing $\dcb$, respectively, and are color-coordinated to reflect best-fits of the $1\n$ model (red) and $3\n$ model (blue).}
    \label{fig:maglim}
\end{figure}

\begin{figure}
    \centering
    \includegraphics[width =0.7\textwidth]{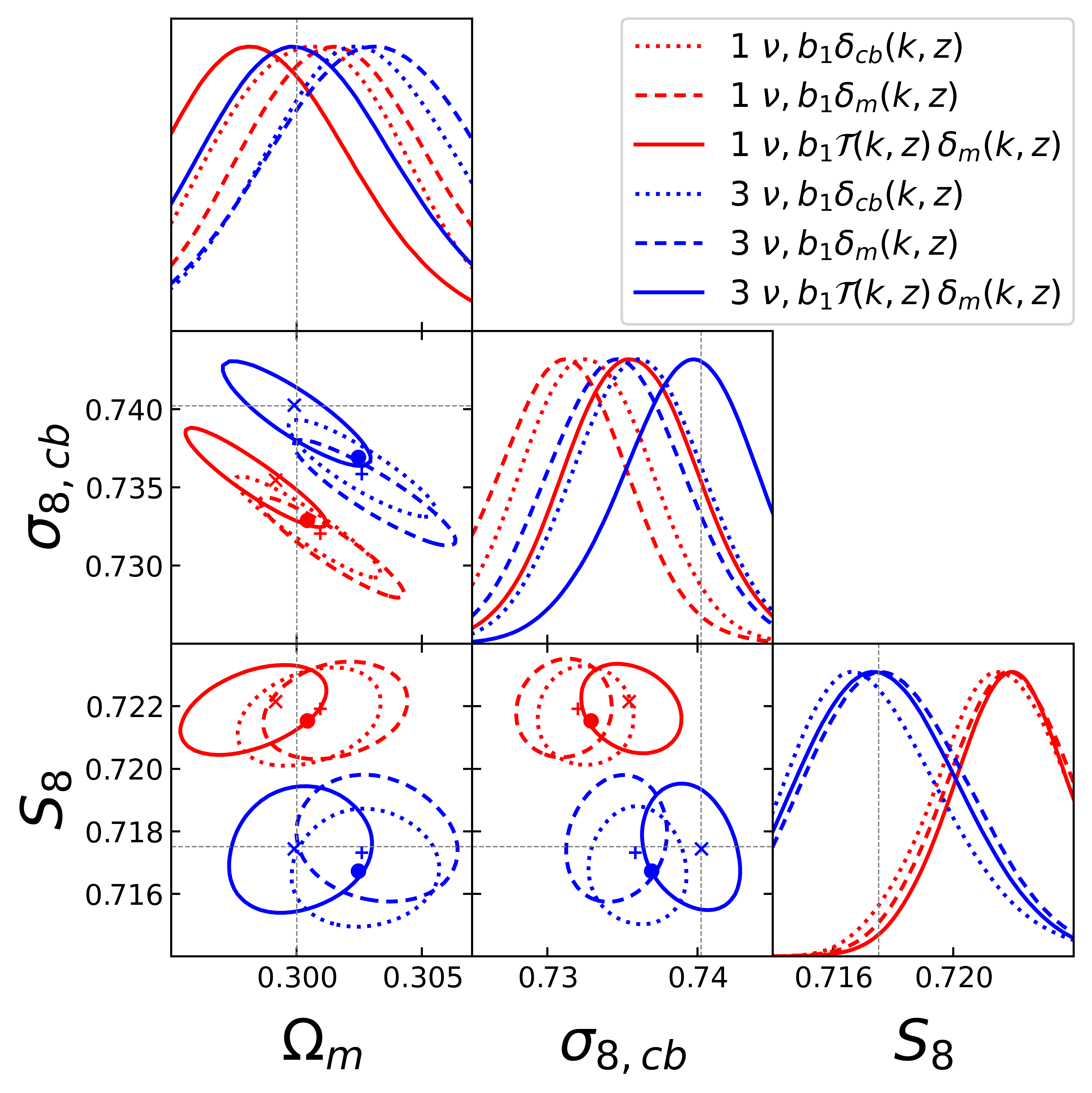}
    \caption{LSSTY1 simulated $0.3\sigma$ constraints of the marginalized 2D parameter contours for the neutrino mass and galaxy bias models listed in the given legend. Best-fit MAP values depicted as cross, plus, and circle points for the NISDB, the constant linear galaxy bias model tracing $\dm$, and the constant linear galaxy bias model tracing $\dcb$, respectively, and are color-coordinated to reflect best-fits of the $1\n$ model (red) and $3\n$ model (blue).}
    \label{fig:lsst}
\end{figure}
 
\subsection{DESY3 Simulated Results}
\label{sec::des}

Figure \ref{fig:maglim} presents the inferred $0.3\sigma$ cosmological parameter constraints of simulated DESY3 analyses. Analyses using the $3 \n$ mass model are depicted in blue and the $1\n$ mass model in red, corresponding to the comparisons in table \ref{tab::neutrino-models}. 

Relative to the scale-independent linear galaxy bias models, parameter contours are 10\% tighter when fitting the fiducial model to the fiducial datavector. Projected parameter shifts of analyses with $3 \n$ are greatest for model (1) (see table \ref{tab::neutrino-models} yet pass our systematic bias criteria. 

Analyses using the $1 \nu$ mass model provide slightly tighter constraints at the expense of inducing parameter biases in cosmological inference. The $1\n$ model with the overall smallest cosmological parameter biases is model (3) (see table \ref{tab::neutrino-models}), which incorporates the NISDB of eq. \eqref{eqn::approx-bias}. The MAP in the $\sigma_{8,\cb}$-$\sigma_8$ projected 2D parameter space is shifted by 2.5--2.9 $\sigma$ while $S_8$ and $\sum m_\n$ projected 1D parameters shift by 0.3--1.1 $\sigma$. The large discrepancy in the $\sigma_{8,\cb}$-$\sigma_8$ plane is induced by assuming a neutrino mass model which does not match that of the fiducial datavector. For the masses and redshifts considered in our analyses, neutrinos are free-streaming at 8 Mpc/h scales, which corresponds to the DESY3 scale-cut in galaxy clustering and is above the scale-cut region in galaxy-galaxy lensing \citep{y3-modeling}. However, the $3\n$ model neutrinos free stream at larger physical scales and suppress the total matter power spectrum at larger physical scales relative to the $1\n$ model. This fact leads to $\sigma_{8}$ being slightly larger in $1\n$ models than $3\n$ models with $\sigma_{8,\cb}$ being relatively unaffected. Further, the free-streaming scale of massive neutrinos changes more rapidly when varying the sum of the neutrino masses in the $1\n$ model compared to the $3\n$ model, resulting in a tighter correlation in the $\sigma_{8,\cb}$-$\sigma_8$ plane. Discrepancies in this plane may be useful in determining systematic biases in cosmological parameter impact tests.

\subsection{LSSTY1 Simulated Results}
\label{sec::lsst}
Figure \ref{fig:lsst} shows the LSSTY1 simulated $0.3 \sigma$ inferred parameter constraints of the $3\n$ and $1\n$ neutrino mass models with projected 1D and 2D parameter shifts in tables \ref{tab:1dparams} and \ref{tab:2dparam}, respectively. We generally observe that inferences made with the fiducial model are up to 20\% more constraining relative to the scale-independent galaxy bias models in $\Omega_\m$, $h_0$, $\sigma_{8}$, and $\sigma_{8,\cb}$. For all models differing from the fiducial model, our adopted systematic bias criteria are exceeded. As shown in table \ref{tab:1dparams}, nearly all model parameters exhibit projected 1D parameter shifts between 0.5--2.3$\sigma$. Projected 2D parameter shifts for $1 \n$ models are at the 1.5--2.3$\sigma$ level in the $S_8$-$\Omega_\m$ plane and at the 8--10$\sigma$ level in the $\sigma_{8,\cb}$-$\sigma_8$ plane.

For the $3 \n$ models, shifts in the $S_8$-$\Omega_\m$ plane are barely within our systematic bias criteria and the 3x2pt $\Delta \chi^2$ is much less than 1 (see table \ref{tab:dchisq}). All non-fiducial models recover biased 1D $\sigma_{8,\cb}$ constraints, whereas 1D constraints in $S_8$ are insensitive to the choice of galaxy bias for a given neutrino mass model; we find interpreting model consistency in quantities directly related to matter clustering (e.g. $\sigma_{8,\mathrm{cb}}$) better reflect systematic biases due to galaxy clustering modeling choices than $S_8$. 

\section{Conclusions}
\label{sec::conclusion}

In this work, we implemented an approximate description of the neutrino-induced scale-dependent bias (NISDB) for use in LSS survey likelihood analyses. We design analyses to determine the impact of linear galaxy bias and neutrino mass model choices on cosmological parameter inference when the fiducial data contains a substantial NISDB. Our approximation, built off the work of \cite{LoVerde_bias}, does not increase the model complexity of linear galaxy bias models and characterizes the small-scale damping of galaxy clustering statistics due to massive neutrinos, providing improvements on constraining cosmological parameters for a flat $\Lambda$CDM universe with massive neutrinos. 

We model a DESY3 and LSSTY1 analysis using noiseless synthetic data that includes a NISDB for a three degenerate-mass neutrino mass model (denoted 3$\n$) at $\sum m_\n = 0.5$eV. For these data vectors, we infer cosmological parameters and different choices of neutrino mass and galaxy bias models as detailed in table \ref{tab::neutrino-models}. We determine whether an analysis is systematically biased by requiring ``unbiased fits'' to yield a 3x2pt $\Delta \chi^2 < 1$ and a shift in the $S_8$-$\Omega_\m$ plane that is less than 0.3$\sigma$ (similar to \citep{y3-modeling}). We additionally investigate parameter shifts in the inferred linear fluctuations of the CDM and baryon field (denoted as $\cb$) at 8 Mpc/h scales as the parameter $\sigma_{8,\cb}$. As neutrinos are free-streaming at 8 Mpc/h scales for the masses in our prior range, shifts in $\sigma_{8,\cb}$ indicate how an assumed neutrino mass or linear galaxy bias model affects matter which clusters at 8 Mpc/h scales. 

We find DESY3 cosmological inferences are not significantly biased by the choice of galaxy bias model, so long as the modeled neutrino mass model corresponds to the $3\n$ model of the fiducial synthetic datavector. However, we find a gain in constraining power of up to 10\% in $\Omega_\m$, $\sigma_{8}$, and $\sigma_{8,\cb}$ when incorporating the NISDB correction. This is not the case for DESY3 inferences utilizing the $1\n$ neutrino mass model, which infers a lower total neutrino mass biased up to 1.1$\sigma$ and induces 2.5--2.9$\sigma$ shifts in the $\sigma_8$-$\sigma_{8,\cb}$ plane. 

With LSSTY1 precision, implementing accurate neutrino mass and galaxy bias models becomes more important: all non-fiducial models result in parameter biases from the fiducial $3\n$ $\sum m_\n = 0.5$eV cosmology. Additionally, inferences using the fiducial NISDB model provide up to 20\% tighter cosmological parameter constraints. Despite modeling the fiducial datavector with a relatively high total neutrino mass, the analyses show the importance of accurate modeling of neutrino-induced effects on galaxy clustering for future 3x2pt analyses, e.g. for LSSTY10. For LSSTY1 analyses which infer cosmological parameters assuming the fiducial three degenerate-mass neutrino mass hierarchy, the systematic bias criteria indicate unbiased cosmological inference despite 0.5--1.0$\sigma$ shifts in the 1D marginalized posterior distributions of $\Omega_\m$, $h_0$, $\sigma_{8}$, and $\sigma_{8,\cb}$. For future surveys, e.g. LSSTY10, additional model consistency tests that are more sensitive to inaccuracies in galaxy clustering models should be investigated to avoid biases in the cosmological inference.

\acknowledgments
We thank Jessie Muir, Arka Banerjee, and Julia Stadler for their helpful discussion and comments. PR and EK are supported in part by Department of Energy grant DE-SC0020247, the David and Lucile Packard Foundation, and an Alfred P. Sloan Research Fellowship. Calculations in this paper used High Performance Computing (HPC) resources supported by the University of Arizona TRIF, UITS, and RDI and maintained by the UA Research Technologies department. 

\bibliographystyle{JHEP.bst}
\bibliography{references.bib}
\end{document}